\documentclass[conference]{IEEEtran}
\IEEEoverridecommandlockouts
\usepackage{amsmath,amssymb,amsfonts}
\usepackage{algorithmic}
\usepackage{graphicx}
\usepackage{textcomp}
\usepackage{xcolor}
\usepackage[sorting=none,style=numeric]{biblatex}
\usepackage{afterpage}
\def\BibTeX{{\rm B\kern-.05em{\sc i\kern-.025em b}\kern-.08em
    T\kern-.1667em\lower.7ex\hbox{E}\kern-.125emX}}
\begin{document}

\title{Towards Robust Real-time Audio-Visual Speech Enhancement\\
\thanks{This work is supported by the UK Engineering and Physical Sciences Research Council (EPSRC) programme grant: COG-MHEAR (Grant reference EP/T021063/1).}
\thanks{Corresponding author: m.gogate@napier.ac.uk}
}
\author{\IEEEauthorblockN{Mandar Gogate, Kia Dashtipour, Amir Hussain}
\IEEEauthorblockA{\textit{School of computing, Edinburgh Napier University} \\
Edinburgh, United Kingdom \\
}
}

\maketitle

\begin{abstract}

The human brain contextually exploits heterogeneous sensory information to efficiently perform cognitive tasks including vision and hearing. For example, during the cocktail party situation, the human auditory cortex contextually integrates audio-visual (AV) cues in order to better perceive speech. Recent studies have shown that AV speech enhancement (SE) models can significantly improve speech quality and intelligibility in very low signal to noise ratio (SNR) environments as compared to audio-only SE models. However, despite significant research in the area of AV SE, development of real-time processing models with low latency remains a formidable technical challenge. In this paper, we present a novel framework for low latency speaker-independent AV SE that can generalise on a range of visual and acoustic noises. In particular, a generative adversarial networks (GAN) is proposed to address the practical issue of visual imperfections in AV SE. In addition, we propose a deep neural network based real-time AV SE model that takes into account the cleaned visual speech output from GAN to deliver more robust SE. The proposed framework is evaluated on synthetic and real noisy AV corpora using objective speech quality and intelligibility metrics and subjective listing tests. Comparative simulation results show that our real time AV SE framework outperforms state-of-the-art SE approaches, including recent DNN based SE models. 
 
\end{abstract}

\begin{IEEEkeywords}
audio-visual, speech enhancement, generative adversarial network
\end{IEEEkeywords}

\section{Introduction}
More than 430 million people worldwide currently suffer from hearing loss. These numbers are expected to reach 2.5 billion by 2050~\cite{world2021hearing}. The most common type of hearing loss is neither curable nor reversible. Studies have shown that the hearing impaired listeners often find themselves in social isolation leading to depression. Hearing aids and cochlear implants are among the most common devices used to compensate for hearing loss. However, even the sophisticated hearing aids that use state-of-the-art speech enhancement (SE) algorithms pose significant problems for the people with hearing loss as these listening device often amplify sounds but do not restore intelligibility in busy social situations~\cite{lesica2018hearing}. Normal hearing listeners in such environments use the audio-visual (AV) nature of speech to suppress background noise and focus on the desired speech. Therefore, researchers have proposed AV SE methods that go beyond traditional audio-only (A-only) SE. 

Speech enhancement, aims to separate speech from background noise, has had a huge impact in recent years due to its applications in automatic hearing aids, cochlear implants, speech recognition, mobile communication, and voice activity detection~\cite{donahue2018exploring}.  Despite extensive research advances in speech enhancement, hearing scenarios are becoming more complex with a wide range of non-stationary acoustic noises, visual speech noises and reverberations in physical space.

\begin{figure}[!t]
    \centering
    \includegraphics[width=\linewidth]{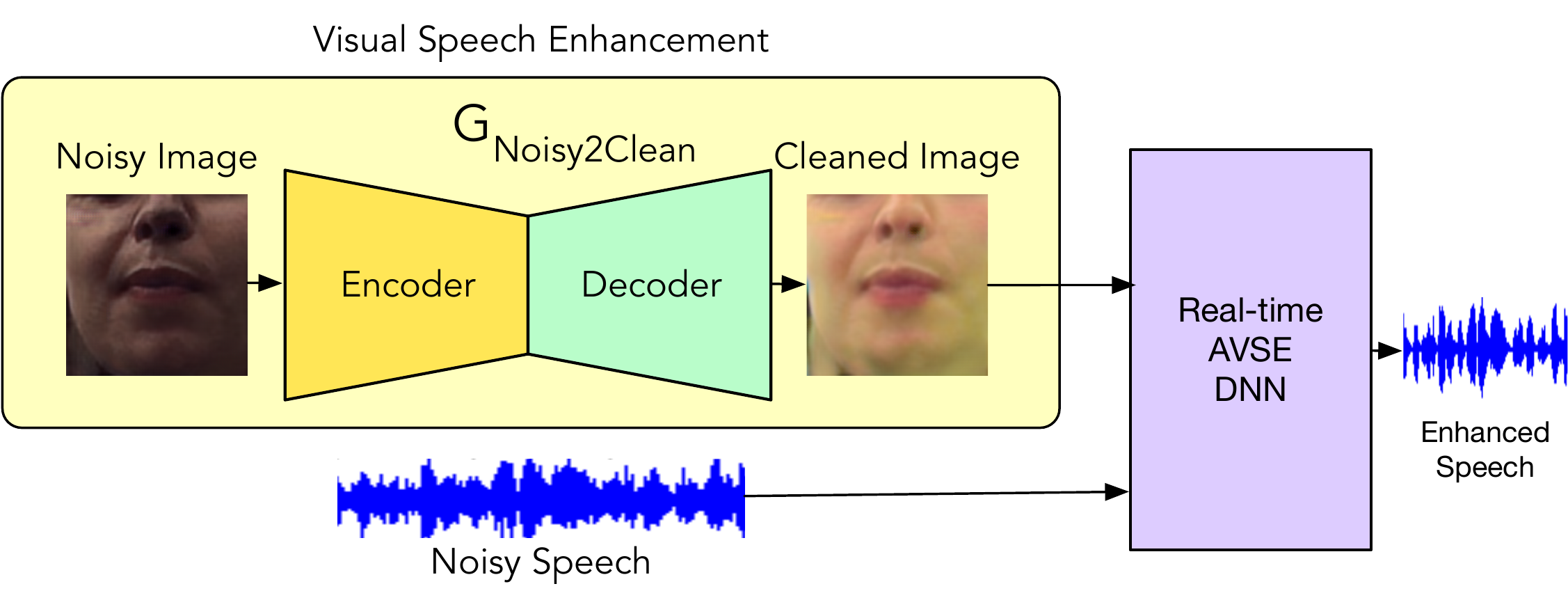}
    \caption{Proposed Real-time Audio-Visual Speech enhancement Framework }
    \label{fig:overview_gan}
\end{figure}

In the literature, extensive studies have been carried out to develop AV SE methods in time-domain and frequency domain~\cite{abel2014novel,afouras2018conversation,hou2018audio,lu2019audio,adeel2020contextual,afouras2019my,michelsanti2019deep,aldeneh2020self,gogate2020cochleanet,kim2021multi,arriandiaga2021audio,gao2021visualvoice,montesinos2021cappella, gogate2020deep}. However, despite significant research in the area of AV SE, real-time processing models, with no or low latency (8-12 ms) remains a formidable technical challenge. Most of the aforementioned methods are computationally complex and are not suitable for processing streaming data with real-time constraints. The processing latency is of particular importance to the hearing impaired listeners because the delayed processing will result in an echo effect and unsynchronised AV cues leading to poor speech intelligibility. 

The aforementioned state-of-the-art AV SE methods employ AV speech corpora recorded in an ideal (e.g studio-like) environment for training and evaluation. However, in real world environments visual speech is often degraded by poor lighting, occlusions and head movements. The imperfections in visual speech degrade the overall performance of AV SE model as compared to A-only SE models in term of subjective speech quality and intelligibility.

In this study, we propose robust real time AV SE framework that is computationally less complex and can process streaming data with low latency. The real-time AV SE framework consists of visual imperfection generative adversarial network (GAN) and real-time AV SE model. In particular, we propose a visual noise robust GAN architecture to enhance the visual imperfections in noisy speech images. The developed GAN ingests the noisy lip images and output an enhanced version by eliminating visual speech noises. The real-time AV SE model ingests the enhanced visual images predicted by the GAN architecture for more robust SE with variety of visual and acoustic noises. To the best of our knowledge, our paper is first to propose a framework that jointly enhance the visual and acoustic speech signal for robust AV SE. The comparative simulation results in terms of objective speech quality and intelligibility metrics (including PESQ, STOI and SI-SDR) and subjective listening tests (using real noisy ASPIRE~\cite{gogate2020cochleanet} and VISION~\cite{gogate2020vision} corpora) show significant performance improvement of the proposed framework as compared to state-of-the-art approaches. Figure~\ref{fig:overview_gan} depicts a top-level overview of our proposed framework.

In summary, this paper presents three major contributions:
\begin{enumerate}
    \item  A generative adversarial network architecture is proposed to address the main limitation of visual imperfection in audio-visual speech enhancement models. To the best of our knowledge, our paper is first to introduce a framework that take into account both acoustic and visual noise for real-time speech enhancement.
    \item A real-time audio-visual speech enhancement model is proposed for low-latency inference on streaming audio-visual data. We present the computational latency of individual blocks in the proposed framework and show the effectiveness of the proposed model for streaming speech enhancement. 
    \item Extensive evaluation using objective metrics and subjective mean opinion score (MOS) listening tests of the proposed framework has been carried out using real noisy ASPIRE and VISION corpora with state-of-the-art and DNN based audio-only and audio-visual speech enhancement approaches including spectral subtraction (SS), Linear minimum mean square error (LMMSE), Speech Enhancement Generative Adversarial Network (SEGAN) and CochleaNet.
\end{enumerate}

The rest of the paper is organised as follows:
Section \ref{sec:relatedwork} presents an overview of audio-visual speech enhancement models proposed in the literature. Section \ref{sec:gan} presents robust feature extraction using generative adversarial network. Section \ref{sec:ssframework} introduces real-time audio-visual speech enhancement model. Section \ref{sec:experiment} presents the experimental results. Finally, section \ref{sec:conclusiongan} concludes the work and present possible future research directions.

\section{Related Work}\label{sec:relatedwork}
In this section, we review related work in the area of AV SE. 

Abel et al.~\cite{abel2014novel} developed two stage AV SE model which uses A-only beamforming and visually derived wiener filter for SE. The model was evaluated using studio recorded GRID corpus with mixture of aircraft and clapping noises. The experimental results demonstrate that the proposed AV model can outperform traditional A-only beamforming approaches. Recently, Afouras et al.~\cite{afouras2018conversation} presented a deep neural network model to separate speaker’s voice using lip region features. The model is trained using studio-quality LRS2 and VoxCeleb2 dataset to predict magnitude and phase of the target signal. Similarly, Gogate et al.~\cite{gogate2018dnn} proposed a speaker independent AV SE model based on deep neural networks. The model is trained and evaluated using synthetic GRID-CHIME3 dataset. However, the main limitation is that the model is evaluated on limited vocabulary Grid corpus.  On the other hand, Hou et al.~\cite{hou2018audio} proposed a deep denoising autoencoder based on convolutional neural network (AVDCNN) for speech enhancement, which combines both audio and visual modalities. Comparative simulation results show that the proposed AVDCNN outperforms state-of-the-art A-only approaches including logMMSE. 

Lu et al.~\cite{lu2019audio} introduced a speaker-independent speech separation model based on AV deep clustering. The model learns time-frequency (T-F) embeddings for AV speech features. The model was trained using GRID and TCD-TIMIT corpora. The experimental results demonstrates that the proposed AV model outperform A-only deep clustering and other state-of-the-art approaches. In addition,
Afouras et al.~\cite{afouras2019my} presented a technique based on the deep learning for AV speech separation to separate speaker’s voice by conditioning on speaker’s lip movement and their voice. The model is trained and evaluated using LRS3 dataset and the experimental results shown the robustness of the model when visual information is occluded. Furthermore, Michelsanti  et al. \cite{michelsanti2019deep} performed set of experiments to understand the impact of Lombard effect on A-only and AV SE. The empirical results indicated the benefits of training system with Lombard AV Grid corpus on speech quality and intelligibility in low SNR environments.

Recently, Adeel et al. \cite{adeel2020contextual,adeel2017towards,adeel2019lip} proposed an enhanced visually derived wiener filter for two-stage AV SE. In the first stage, the model extracts lip reading features using a deep neural network based regression model. The extracted lip reading features are then fed to an enhanced visually derived wiener filter for SE. The model is trained and evaluated using GRID corpus combined with CHiMe3 noises. The model contextually integrates AV features for robust SE. 
In addition, Gogate et al. \cite {gogate2020cochleanet} proposed speaker independent AV deep neural network for ideal binary mask estimation to remove the noise from the speech. The model is trained using GRID corpus and evaluated using a range of speaker, noise, and language independent test sets. 

\begin{figure*}[!t]
    \centering
    \includegraphics[width=0.8\linewidth]{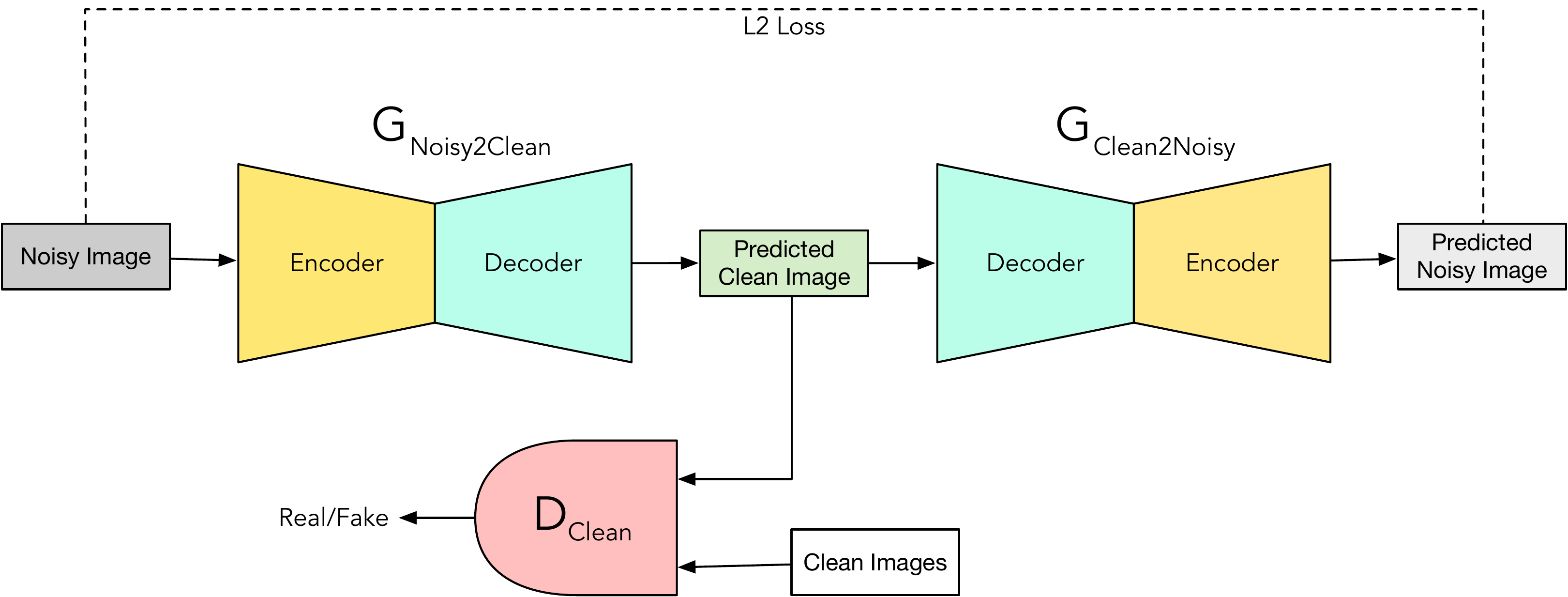}
    \caption{Proposed visual speech enhancement framework}
    \label{fig:overview_gan_framework}
\end{figure*}

More recently, Arriandiaga et al.~\cite{arriandiaga2021audio} proposed a method based for SE in multi-talker conversation environment. The facial landmarks are used as an alternative to visual lip images which is widely used in AV SE literature. The model is trained using GRID corpus to predict ideal amplitude mask in order to filter noisy audio. The experiments show that the method achieved better performance with low latency and computational cost. Additionally, Gao et al.~\cite{gao2021visualvoice} developed method to learn cross-modal speaker embeddings and SE in the multi-task setting. The model integrates lip motion features, facial attribute embedding for robust SE.  The LRS2, VoxCeleb1 and VoxCeleb2 corpora are used to measure the performance of the approach. The results display that the approach can generalise well in real-world scenarios. Finally, Montesinos et al.~\cite{montesinos2021cappella} proposed AV convolutional network based on graphs to separate singing voice by exploiting acoustic and visual information. The experimental results show the proposed approach outperforms state-of-the-art A-only singing voice separation approaches.

In summary, it can be seen that most of the aforementioned AV SE models are computationally complex and non-causal and hence cannot be used for real-time SE. In addition, the models are trained using corpora recorded in an ideal (studio-like) environment e.g. GRID, TCD-TIMIT. This limits the ability of AV SE models to generalise in real noisy environments where both visual and acoustic speech is mixed with a range of noises.

\section{Robust visual feature extraction using Generative Adversarial Networks}\label{sec:gan}

Generative Adversarial Network (GAN) obtained extraordinary performance in wide range of real world applications including image translation, image segmentation, art generation etc. In order to train the model, we required set of input and output images for training. The collection of such parallel image corpus is infeasible in addressing the issue of imperfect visuals in AV SE. This limitation can be addressed by a modified GAN architecture similar to CycleGAN, which exploits cycle consistency loss to enable training without the need for paired data. The model can translate from one domain to another using unpaired samples from individual domains. The proposed GAN, shown in Fig. \ref{fig:overview_gan_framework}, often consists of two generator and discriminator networks. The first generator maps the input image from domain A to B and the second generator maps from domain B to A. The two discriminator is used to estimates the distance between the predicted samples and actual samples from each domain for model training. The domain A to B generator can be used separately for the task of image translation once the training is complete. 

In this section, we exploit the aforementioned GAN architecture for robust visual speech denoising. The model learns the mapping between noisy visual speech consisting of improper lighting, exposure and contract and clean visual speech. The predicted visual speech is then fed to a real-time AV model for more robust SE.

\begin{figure*}[!t]
    \centering
    \includegraphics[width=\linewidth]{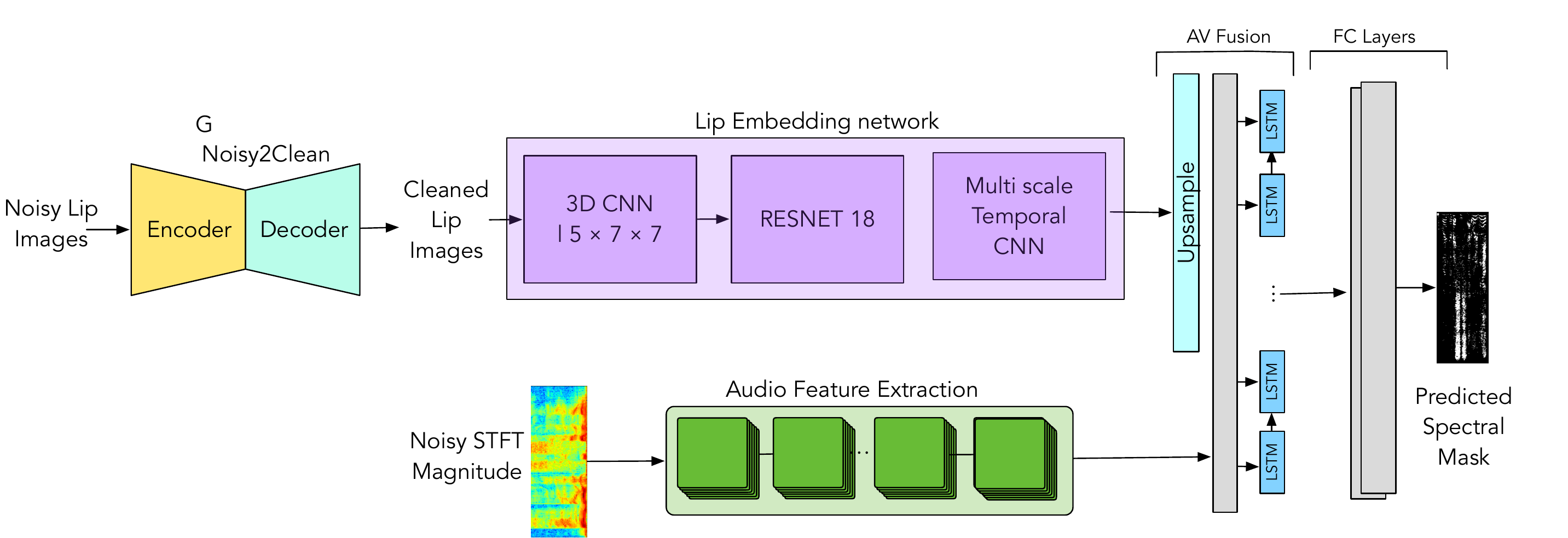}
    \caption{Proposed Real-time Audio-Visual Speech enhancement Framework}
    \label{fig:CochleaNet_GAN}
\end{figure*}

\subsection{Data Representation}

\paragraph{Input features} The GAN framework uses noisy cropped lip region input. The cropped 96 x 96 lip region is extracted from from real noisy VISION~\cite{gogate2020vision} and were used as noisy image samples. 

\paragraph{Output} The GAN framework generates the clean cropped lip region. It is worth to mention that, the Grid corpus is recorded in the studio environment and it can be used without any modification as training set.

\subsection{Network Architecture}

We adopt the generator architecture from Zhu et al. \cite{zhu2017unpaired} that achieve state-of-the-art results for style transfer and image-translation tasks. The generator network consists of three basic building blocks: (1) encoder (2) transformer (3) decoder. The encoder block consists of three convolutional layers that mitigates the illustration by one fourth of actual image size. Encoder output is ingested to a transformer block consisting of residual blocks. The decoder block consists of 3 deconvolutional layers to regenerate the original size of input image. For the discriminator we use the 70 x 70 pixel PatchGAN \cite{isola2017image} architecture which able to classify if 70 x 70 overlapping image patches are real or fake. This majorly reduces the number of parameters as compared to full image discriminator.

\section{Real-time Audio-visual Speech enhancement Framework} \label{sec:ssframework}

This section presents the steps involved in end-to-end processing of the proposed framework, shown in Fig.~\ref{fig:CochleaNet_GAN}, to output enhanced speech given AV speech as input. 

\subsection{Data Representation}\label{subsec:video-and-audio-representation_avspeech}
\paragraph{Input features} The deep neural network used both the audio and visual as the input features. Three seconds lip embeddings are used for the batch training and the cropped 96 x 96 lip region is extracted from the video and fed to lip embedding network to generate 75 x 512 dimension vector of lip embeddings for three second of video (assuming 25fps sampling rate). The audio input is divided into windows and a short-time Fourier transform (STFT) of audio segment is calculated. The magnitude of STFT is fed to the models as noisy input. The model is trained on 3 second segments and can be used for inference of arbitrary lengths of noisy video.

\paragraph{Output} 
The ideal binary mask is used as the output of the network. IBM is a multiplicative spectrogram mask that shows a the time-frequency (T-F) relationship between the source audio and interfering noise. The IBM has a value of 1 where the local SNR is higher than local criterion (LC) and zero otherwise. The LC is calculated using the source audio and interfering noise. As the interfering noise is unavailable in real noisy environment the mask cannot be calculated using the aforementioned definition. However, the IBM can be predicted using data-driven machine learning models using synthetic noisy datasets that learn the correlation between noisy speech, visual lip images and the binary mask.

\subsection{Network Architecture}\label{subsec:network-architecture-avspeech}

A detailed outline of the proposed framework is shown in Figure~\ref{fig:CochleaNet_GAN}. The individual components are explained in the subsequent sections. 

\subsubsection{Noisy audio feature extraction}
The number of convolution filters, strides and dilation used in the acoustic feature extraction is detailed in Table~\ref{tab:audFeatEx_gan}.

\begin{figure*}[!t]
    \centering
    \includegraphics[width=\textwidth]{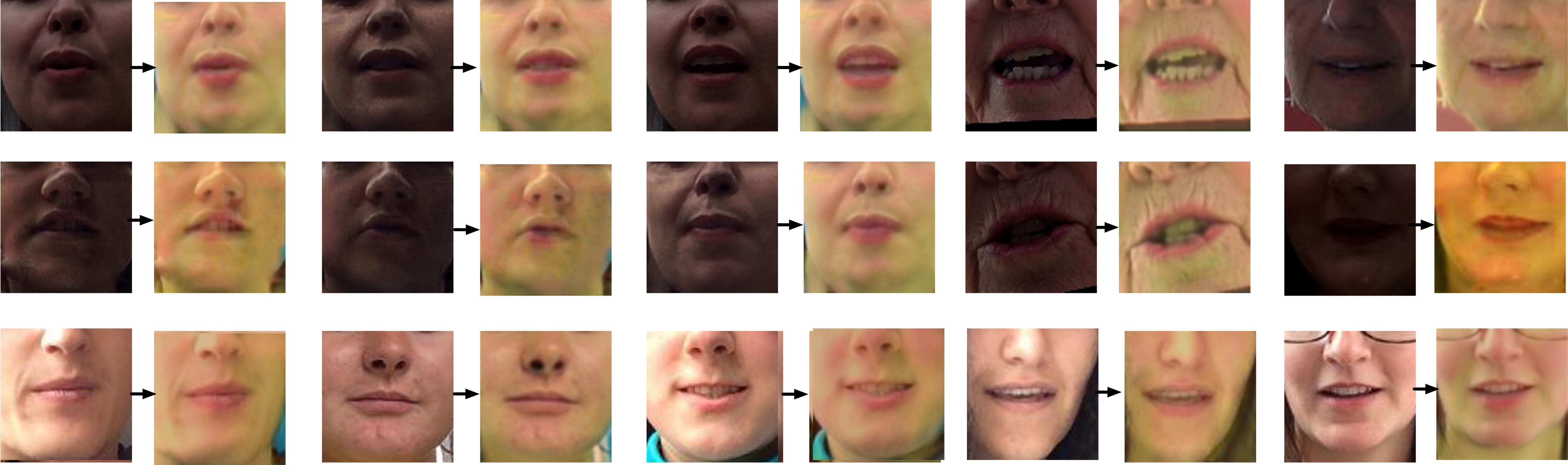}
    \caption{Visual Speech Enhancement Evaluation on Unseen speakers from VISION Corpus}
    \label{fig:gan_vision_eval}
\end{figure*}

\subsubsection{Lip Embedding Network}
The lip embedding network shown in Fig.~\ref{fig:CochleaNet_GAN} consists of a 3D-convolutional network, RESNET-18 and temporal CNN layer. The 3D-CNN consists of single filter with size of 5 x 7 x 7. The 3D-CNN features are fed to standard RESNET-18 architecture \cite{targ2016resnet}. The output of the residual network is fed to a multi-scale temporal CNN as described in \cite{martinez2020lipreading}. The temporal CNN outputs are fed to a fully-connected layer for word classification. The model is trained end-to-end with LRW dataset for lip reading \cite{chung2016}. The fully connected network in the trained model was removed for extracting lip embedding features.  

\subsubsection{Multimodal Fusion}

The sampling rate for visual features are 25 frame per second and 75 vector per second for audio feature. In order to match the audio STFT sampling rate (75 frames/sec), the visual frames (sampled at 25 frames/sec) are upsampled using the repetition of the element three times in the temporal dimension~\cite{gogate2017novel}.

The acoustic features and visual features are combined using the time dimension and fed to the LSTM layer with 622 units. 
The two fully connected layers consisted LSTM outputs along with 622 neurons and ReLU activation function. It is to be noted that, the fully connected layer weights are shared across time dimension. The extracted feature are fed into fully connected layers with 622 neurons and sigmoid activation function.

\begin{table}[!t]

\centering
\caption{Noisy Audio Feature Extraction}
\begin{tabular}{lccccccccc}
\hline

 & conv1 & conv2 & conv3 & conv4 & conv5 \\ \hline
Num filters & 64 & 64 & 64 & 64 & 4 \\ 
Filter size & 5 x 5 & 5 x 5 & 5 x 5 & 5 x 5 & 1 x 1 \\
Dilation & 1 x 1 & 1 x 1 & 1 x 1 & 1 x 1 & 1 x 1 \\ \hline
\label{tab:audFeatEx_gan}

\end{tabular}
\end{table}

\subsection{Speech Resynthesis}\label{subsec:resynthesis_gan}

The model predicts the time-frequency binary mask when the noisy spectrogram and lip embeddings are fed into the network. The enhanced speech is obtained by multiplying the predicted magnitude mask with noisy magnitude spectrogram and inverse STFT. It is to be noted that, the phase of noisy signal and masked magnitude is used to resynthesize enhanced speech.

\section{Experimental Results} \label{sec:experiment} 
We evaluated the performance of our approach using state-of-the-art A-only and AV approaches including CochleaNet in range of synthetic and real noisy scenarios.

\subsection{Synthetic Dataset}

The synthetic AV Grid corpus~\cite{Grid} is used for the training and evaluation of the model. All the thirty-three speakers with one thousand utterances for each speaker used. The Grid utterances are randomly mixed with the noises from CHiME 3~\cite{CHiME3}, which compromise different types of noises including bus, streets, cafeteria at SNR ranging from -12 to 9 dB with step size of 3dB. 
In total, the training, evaluation and test set consists of 25000, 4000 and 4000 utterances respectively.  It is to be noted that, there is no overlap of speakers between train, test and validation set. In addition, the Grid corpus consist of limited vocabulary.

\begin{table*}[!t]
\centering
\caption{Objective Evaluation using PESQ scores for speaker independent test set (GRID - CHIME3)}
\label{tab:pesqgc3_gan}

\begin{tabular}{lcccccccc}
\hline
\textbf{dB} & -12 & -9 & -6 & -3 & 0 & 3 & 6 & 9 \\
 \hline           
Noisy & 1.31 & 1.40 & 1.55 & 1.70 & 1.88 & 2.08 & 2.28 & 2.46 \\
 \hline
SS & 1.13 & 1.23 & 1.41 & 1.60 & 1.83 & 2.09 & 2.35 & 2.58 \\
 \hline
LMMSE & 1.36 & 1.52 & 1.74 & 1.96 & 2.17 & 2.40 & 2.59 & 2.76 \\
 \hline
SEGAN+ & 0.83 & 1.07 & 1.45 & 1.80 & 2.12 & 2.38 & 2.58 & 2.76 \\
 \hline
CochleaNet A & 1.85 & 2.04 & 2.24 & 2.40 & 2.53 & 2.64 & 2.74 & 2.81 \\
 \hline
CochleaNet audio-visual & 1.87 & 2.07 & 2.23 & 2.37 & 2.48 & 2.59 & 2.68 & 2.76 \\
 \hline
A-only & 1.88 & 2.07 & 2.24 & 2.37 & 2.54 & 2.63 & 2.75 & 2.82 \\
 \hline
audio-visual & 1.95 & 2.11 & 2.28 & 2.42 & 2.55 & 2.66 & 2.78 & 2.89 \\
 \hline
Oracle IBM & 2.03 & 2.20 & 2.34 & 2.47 & 2.59 & 2.70 & 2.82 & 2.91 \\
\hline
\end{tabular}
\end{table*}

\begin{table*}[!t]
\centering
\caption{Objective Evaluation using PESQ scores for large vocabulary test set (TCD-TIMIT + MUSAN) }
\label{tab:pesqtm_gan}

\begin{tabular}{lcccccccc}
\hline
\textbf{dB} & -12 & -9 & -6 & -3 & 0 & 3 & 6 & 9\\ \hline
Noisy            & 1.48 & 1.56 & 1.62 & 1.69 & 2.23 & 2.33 & 2.44 & 2.50\\ \hline
SS & 1.08 & 1.13 & 1.19 & 1.27 & 1.79 & 1.89 & 2.06 & 2.18\\ \hline
LMMSE & 1.44 & 1.45 & 1.61 & 1.62 & 2.04 & 2.15 & 2.25 & 2.34\\ \hline
SEGAN+ & 1.60 & 1.74 & 1.77 & 1.84 & 2.32 & 2.44 & 2.58 & 2.66\\ \hline
CochleaNet A & 1.81 & 1.90 & 2.02 & 2.13 & 2.37 & 2.47 & 2.52 & 2.56\\ \hline
CochleaNet audio-visual & 1.90 & 2.00 & 2.12 & 2.18 & 2.48 & 2.56 & 2.62 & 2.66\\ \hline
Proposed A-only & 1.92 & 2.03 & 2.13 & 2.44 & 2.50 & 2.53 & 2.57 & 2.60\\ \hline
Proposed audio-visual & 2.05 & 2.17 & 2.27 & 2.60 & 2.67 & 2.70 & 2.74 & 2.78\\ \hline
Oracle IBM & 2.16 & 2.29 & 2.39 & 2.74 & 2.81 & 2.84 & 2.89 & 2.92\\ \hline
\end{tabular}
\end{table*}

  \begin{table*}[!t]
\centering
\caption{Objective Evaluation using PESQ scores for language-independent test set (Hou et al. \cite{hou2018audio} + NOISEX92) }
\label{tab:pesqac_gan}

\begin{tabular}{lcccccccc}
\hline
\textbf{dB} & -12 & -9 & -6 & -3 & 0 & 3 & 6 & 9\\ \hline
Noisy            & 1.04 & 1.25 & 1.29 & 1.31 & 1.40 & 1.49 & 1.55 & 1.61\\ \hline
SS & 0.63 & 1.06 & 0.99 & 0.98 & 1.28 & 1.23 & 1.36 & 1.34\\ \hline
LMMSE & 1.21 & 1.42 & 1.39 & 1.40 & 1.40 & 1.44 & 1.61 & 1.44\\ \hline
SEGAN+ & 1.14 & 1.30 & 1.16 & 1.45 & 1.59 & 1.66 & 1.71 & 1.74\\ \hline
CochleaNet A & 1.28 & 1.42 & 1.56 & 1.53 & 1.66 & 1.72 & 1.79 & 1.74\\ \hline
CochleaNet audio-visual & 1.23 & 1.45 & 1.44 & 1.46 & 1.66 & 1.68 & 1.74 & 1.75\\ \hline
Proposed A-only & 1.32 & 1.53 & 1.59 & 1.61 & 1.70 & 1.73 & 1.79 & 1.78\\ \hline
Proposed audio-visual & 1.44 & 1.57 & 1.65 & 1.58 & 1.70 & 1.75 & 1.79 & 1.81\\ \hline
Oracle IBM & 1.55 & 1.69 & 1.77 & 1.70 & 1.83 & 1.88 & 1.92 & 1.95\\ \hline
\end{tabular}
\end{table*}

The large vocabulary TCD-TIMIT~\cite{TCDTIMIT} corpus is used to understand the performance of the model which consist of 56 speakers with 5488 utterances. Each of the utterances are mixed randomly with the noises from MUSAN~\cite{snyder2015musan}. In addition, for language independet evaluation Mandarin dataset~\cite{hou2018audio} consists of 320 utterances is combined with different types of noise from NOISEX-92~\cite{NOISEX92}.

\subsection{Data preprocessing}

\subsubsection{Audio}

In order to pre-process the audio signals, the 16 khz mono-channel has been used. After resmapling the audio, the signals were segmented into N 78 millisecond frames and also the 17\% increment rate. The hanning window is applied to the frame to generate 622-bin STFT magnitude spectogram.

\subsubsection{Video}

The TCD-TIMIT and Grid corpora are recorded at 25 frame per second (fps). In addition, the Mandarin dataset~\cite{hou2018audio} were recorded at 30 fps, ffmpeg~\cite{newmarch2017ffmpeg} is used to down sampled Mandarin dataset to 25 fps. In order to extract the lip images at 25 fps for the speakers the minified dlib~\cite{dlib09} model was employed. A square region around the lip-centre was extracted using landmark points. The extracted lip region is resized to a square of size 96 pixels.  The cropped lip region is fed into the GAN followed by lip embedding network to extract 512 dimensional embedding for each lip image.

\subsection{Experimental Setup}

The model is developed using TensorFlow library and is trained using NVIDIA 2080Ti GPUs. The speakers present in synthetic GRID + ChiME3 corpus are split into train (77\%), validation (11\%) and test set (11\%) for speaker independent evaluation. In summary, the training, validation and testing set consists of 25000, 4000 and 4000 utterances respectively. It is to be noted that, the speakers were divided to ensure equal gender representation across all sets.  Furthermore, 50 epochs and Adam optimiser with the learning rate (lr) of $3e-4$ are used to train the network . The lr is divided by two when the validation binary cross entropy stops reducing for three consecutive epochs.

\subsection{Qualitative evaluation of Proposed GAN model}

The GAN model is evaluated on the unseen utterances from real Noisy VISION corpus. The real noisy VISION corpus consists of a number of visual imperfections like improper lighting, exposure and occlusions. Fig. \ref{fig:gan_vision_eval} shows some of the unprocessed and processed sample frames from real noisy VISION corpus. It can be observed that, the model considerably enhanced visual imperfections present in the visual speech. We furthermore, envision that the developed model will enable the deployment of AV SE models in wide range of real world environment.

\subsection{Objective evaluation on Synthetic mixtures}

In order to evaluate the quality of the speech, subjective listening test are conducted to ask users to listen to audio and compare the speech quality difference between the processed and unprocessed audio samples. It is worth to mentioning that, conducting subjective listening tests is time consuming as the size of the data increased and the outputs may not show the actual distribution. In such scenarios, PESQ~\cite{pesq}, STOI~\cite{stoi}, and SI-SDR~\cite{sisdr} are used as objective evaluation metrics to approximate subjective listening tests. The proposed model has been compared with the state-of-the-art SE models using the aforementioned evaluation metrics. It is to be noted that, the Grid + ChiME 3, TCD TIMIT + MUSAN and Hou et al~\cite{hou2018audio} + NOISEX-92 are used for speaker independent, large-vocabulary and language independent  evaluation of the proposed framework.

\subsubsection{Perceptual Evaluation of Speech quality (PESQ) comparison} 
Perceptual Evaluation of Speech quality (PESQ)~\cite{pesq} is one the most well-known evaluation metric used to predict the subjective listening test scores in the SE and preliminary results display that correlate well with the subjective listening tests~\cite{hu2007evaluation}. The PESQ scores for proposed A-only SE and AV SE model, A-only CochleaNet, AV CochleaNet~\cite{gogate2020cochleanet}, SEGAN~\cite{pascual2017segan}, SS, and LMMSE for speaker-independent, large-vocabulary and language independent test set is presented in Table~\ref{tab:pesqgc3_gan},~\ref{tab:pesqtm_gan},~\ref{tab:pesqac_gan} respectively.

\begin{table*}[!t]
\centering
\caption{Objective Evaluation using STOI scores for speaker independent test set (GRID - CHIME3)}
\label{tab:STOI_gc3_GAN}

\begin{tabular}{lcccccccc}
\hline
\textbf{dB} & -12 & -9 & -6 & -3 & 0 & 3 & 6 & 9\\ \hline
Noisy            & 0.41 & 0.45 & 0.49 & 0.54 & 0.59 & 0.64 & 0.68 & 0.72\\ \hline
SS & 0.36 & 0.40 & 0.45 & 0.50 & 0.56 & 0.61 & 0.67 & 0.71\\ \hline
LMMSE & 0.39 & 0.43 & 0.48 & 0.53 & 0.58 & 0.63 & 0.68 & 0.71\\ \hline
SEGAN+ & 0.31 & 0.39 & 0.49 & 0.58 & 0.65 & 0.70 & 0.74 & 0.76\\ \hline
CochleaNet A & 0.51 & 0.57 & 0.59 & 0.63 & 0.70 & 0.74 & 0.76 & 0.77\\ \hline
CochleaNet audio-visual & 0.53 & 0.58 & 0.62 & 0.66 & 0.70 & 0.73 & 0.75 & 0.77\\ \hline
Proposed A-only & 0.55 & 0.60 & 0.62 & 0.65 & 0.69 & 0.74 & 0.76 & 0.78\\ \hline
Proposed audio-visual & 0.59 & 0.63 & 0.66 & 0.69 & 0.72 & 0.76 & 0.78 & 0.78\\ \hline
Oracle IBM & 0.61 & 0.65 & 0.68 & 0.71 & 0.74 & 0.77 & 0.79 & 0.80\\ \hline
\end{tabular}
\end{table*}

\begin{table*}[!t]
\centering
\caption{Objective Evaluation using STOI scores for large vocabulary test set (TCD-TIMIT + MUSAN)}
\label{tab:STOI_tcd_GAN}

\begin{tabular}{lcccccccc}
\hline
\textbf{dB} &-12 &-9 &-6 &-3 &0 &3 &6 &9\\ \hline
Noisy            &0.31 &0.34 &0.43 &0.50 &0.60 &0.64 &0.70 &0.73\\ \hline
SS &0.30 &0.35 &0.42 &0.46 &0.59 &0.62 &0.70 &0.73\\ \hline
LMMSE &0.46 &0.48 &0.53 &0.55 &0.66 &0.69 &0.73 &0.76\\ \hline
SEGAN+ &0.44 &0.47 &0.52 &0.55 &0.65 &0.68 &0.73 &0.77\\ \hline
CochleaNet A &0.48 &0.51 &0.54 &0.61 &0.67 &0.70 &0.75 &0.78\\ \hline
CochleaNet audio-visual &0.51 &0.55 &0.60 &0.61 &0.71 &0.73 &0.76 &0.79\\ \hline
Proposed A-only &0.52 &0.56 &0.61 &0.63 &0.72 &0.74 &0.78 &0.80\\ \hline
Proposed audio-visual &0.64 &0.66 &0.68 &0.72 &0.74 &0.75 &0.80 &0.81\\ \hline
Oracle IBM &0.72 &0.74 &0.77 &0.79 &0.81 &0.82 &0.84 &0.85\\ \hline
\end{tabular}
\end{table*}

\begin{table*}[!t]
\centering
\caption{Objective Evaluation using STOI scores for language-independent test set (Hou et al. \cite{hou2018audio} + NOISEX92)}
\label{tab:STOI_avchinese_GAN}

\begin{tabular}{lcccccccc}
\hline
\textbf{dB} &-12 &-9 &-6 &-3 &0 &3 &6 &9\\ \hline
Noisy            &0.54 &0.71 &0.68 &0.73 &0.78 &0.78 &0.85 &0.86\\ \hline
SS &0.42 &0.58 &0.58 &0.62 &0.70 &0.72 &0.77 &0.80\\ \hline
LMMSE &0.52 &0.70 &0.66 &0.71 &0.76 &0.75 &0.83 &0.82\\ \hline
SEGAN+ &0.52 &0.66 &0.58 &0.70 &0.76 &0.76 &0.82 &0.85\\ \hline
CochleaNet A &0.54 &0.73 &0.70 &0.76 &0.81 &0.82 &0.86 &0.88\\ \hline
CochleaNet audio-visual &0.56 &0.73 &0.70 &0.75 &0.81 &0.80 &0.85 &0.87\\ \hline
Proposed A-only &0.58 &0.75 &0.71 &0.77 &0.82 &0.81 &0.87 &0.88\\ \hline
Proposed audio-visual &0.70 &0.80 &0.77 &0.81 &0.86 &0.86 &0.87 &0.88\\ \hline
Oracle IBM &0.81 &0.88 &0.87 &0.90 &0.92 &0.92 &0.94 &0.94\\ \hline
\end{tabular}
\end{table*}

The experimental results shows that our proposed AV model outperform SS~\cite{boll1979spectral}, LMMSE~\cite{ephraim1985speech}, SEGAN+~\cite{pascual2017segan} A-only CochleaNet, and AV CochleaNet~\cite{gogate2020cochleanet}. The proposed AV model is less affected when the visual imperfection are added. 

In addition, AV outperforms A-only model in low SNR particularly $SNR < 0$  dB, mainly where the AV model achieved PESQ score of 1.95 (-12 dB), 2.11 (-9 db), and 2.28 (-6db). Whereas, A-only model achieved PESQ score of 1.88 (-12 dB), 2.07 (-9 db), and 2.24 (-6dB) for speaker independent Grid ChiME 3 test set. 

On the other hand, in the high SNR ($SNR >= 0$ dB) AV performs similar to A-only model. Specifically, AV achieved 2.55, 2.66 and 2.78 PESQ scores for 0dB, 3dB and 6dB respectively. The A-only model achieved PESQ score of 2.54, 2.63, and 2.75 for 0dB, 3dB and 6dB respectively. It can be seen that, the proposed real-time A-only and AV model significantly outperform A-only and AV CochleaNet.

\subsubsection{Short Term Objective Intelligibility (STOI) comparison} STOI is one of the most widely used alternative to PESQ that shows high correlation with subjective listening tests~\cite{stoi}. The STOI scores for proposed A-only SE and AV SE model, SEGAN, SS, LMMSE, A-only and AV CochleaNet for speaker-independent, large-vocabulary and language independent test set is presented in Table~\ref{tab:STOI_gc3_GAN},~\ref{tab:STOI_tcd_GAN},~\ref{tab:STOI_avchinese_GAN} respectively.

\begin{figure*}[!t]
    \centering
    \includegraphics[width=0.8\textwidth]{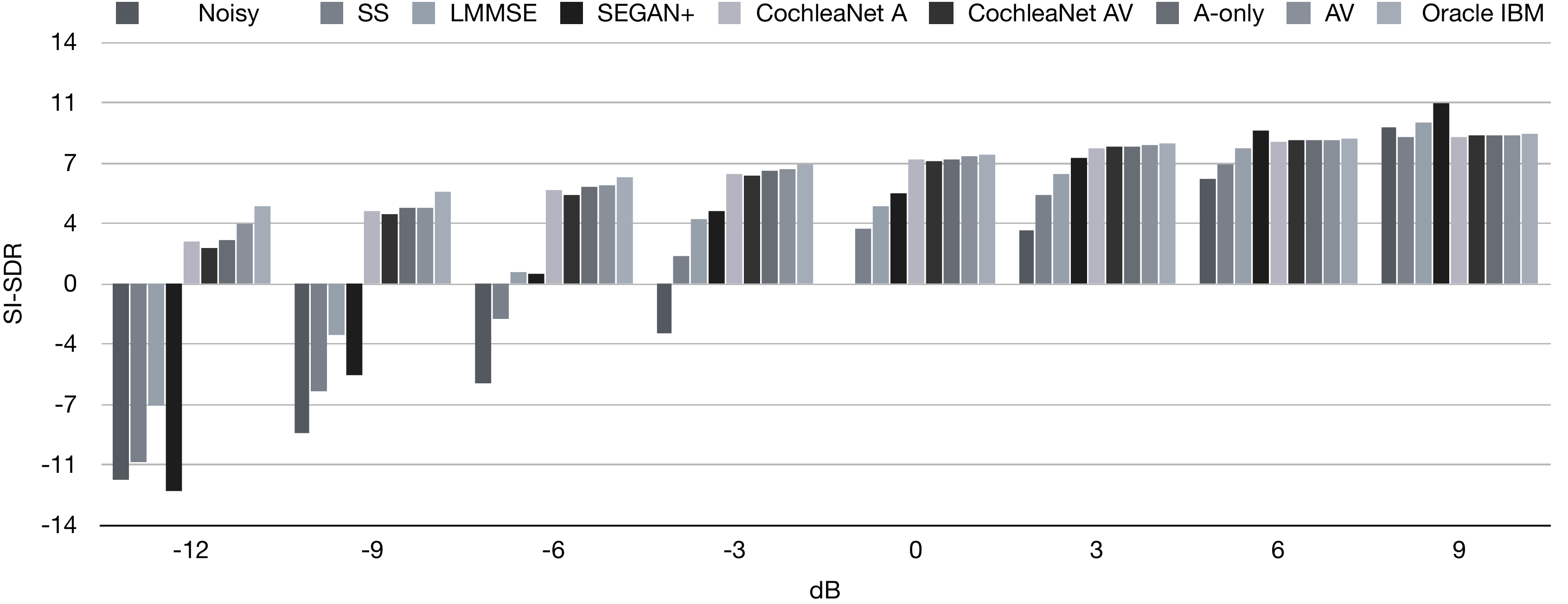}
    \caption{Objective Evaluation using SI-SDR metric for speaker independent test (GRID + CHIME3)}
    \vspace{0.5cm}
    \label{fig:SISDR_gc3_GAN}
        \includegraphics[width=0.8\textwidth]{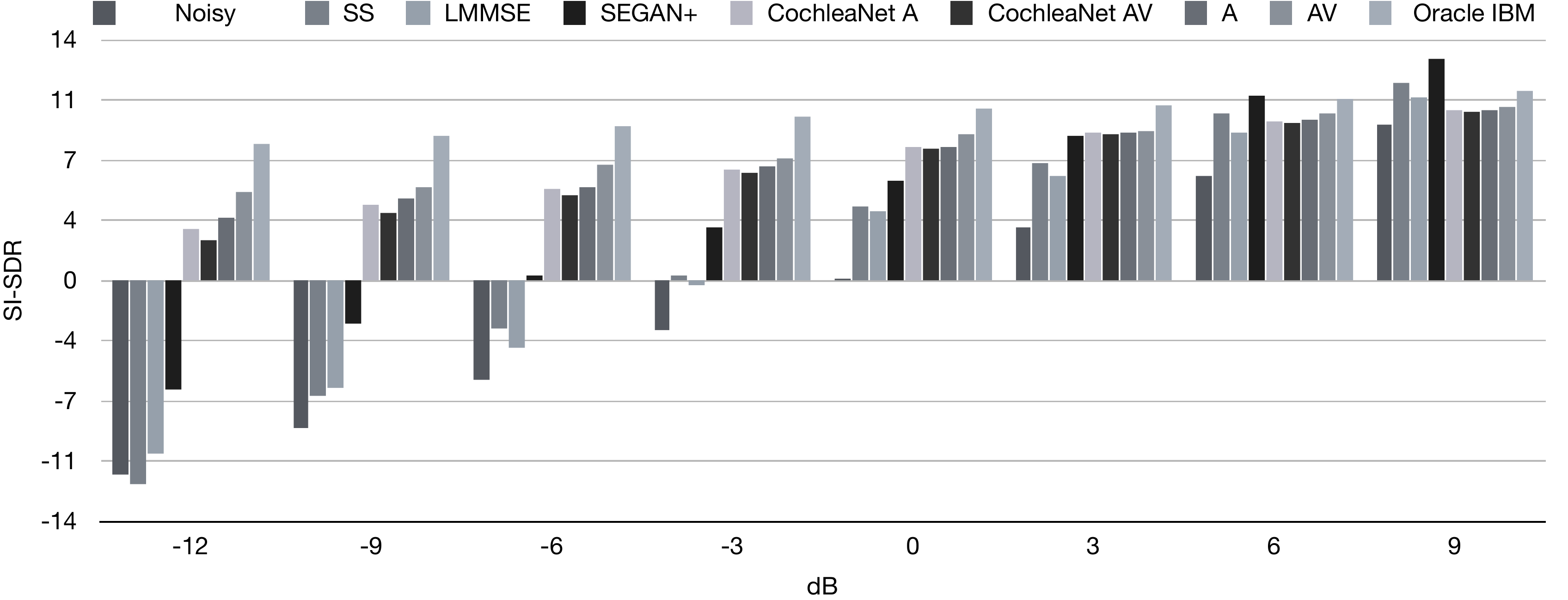}
    \caption{Objective Evaluation using SI-SDR metric for large vocabulary test set (TCD-TIMIT + MUSAN)}
    \vspace{0.5cm}
    \label{fig:SISDR_tcd_GAN}
    \includegraphics[width=0.8\textwidth]{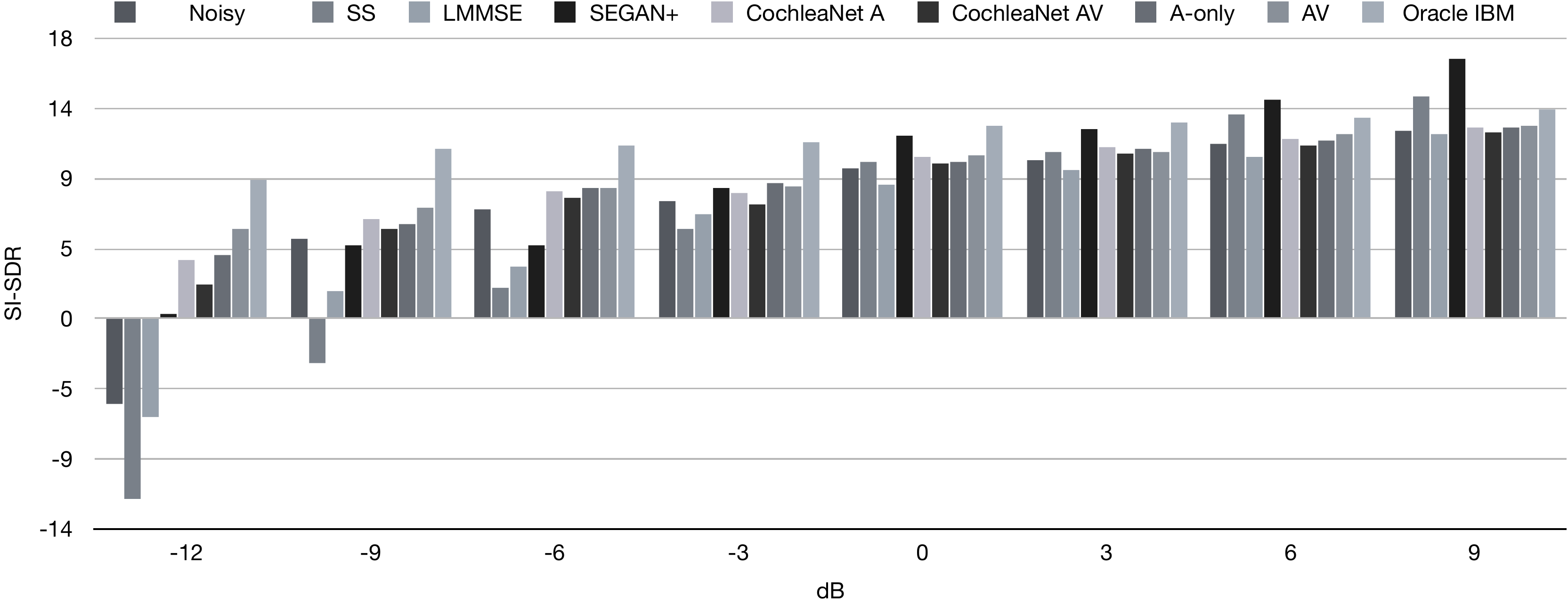}
    \caption{Objective Evaluation using SI-SDR metric for language independent test set (Hou et al. \cite{hou2018audio} + NOISEX-92)}
    \vspace{0.5cm}
    \label{fig:SISDR_hou_GAN}
\end{figure*}

 \begin{figure*}[!t]
    \centering
    \includegraphics[clip, trim=2cm 0.5cm 2cm 0.5cm,width=0.7\linewidth]{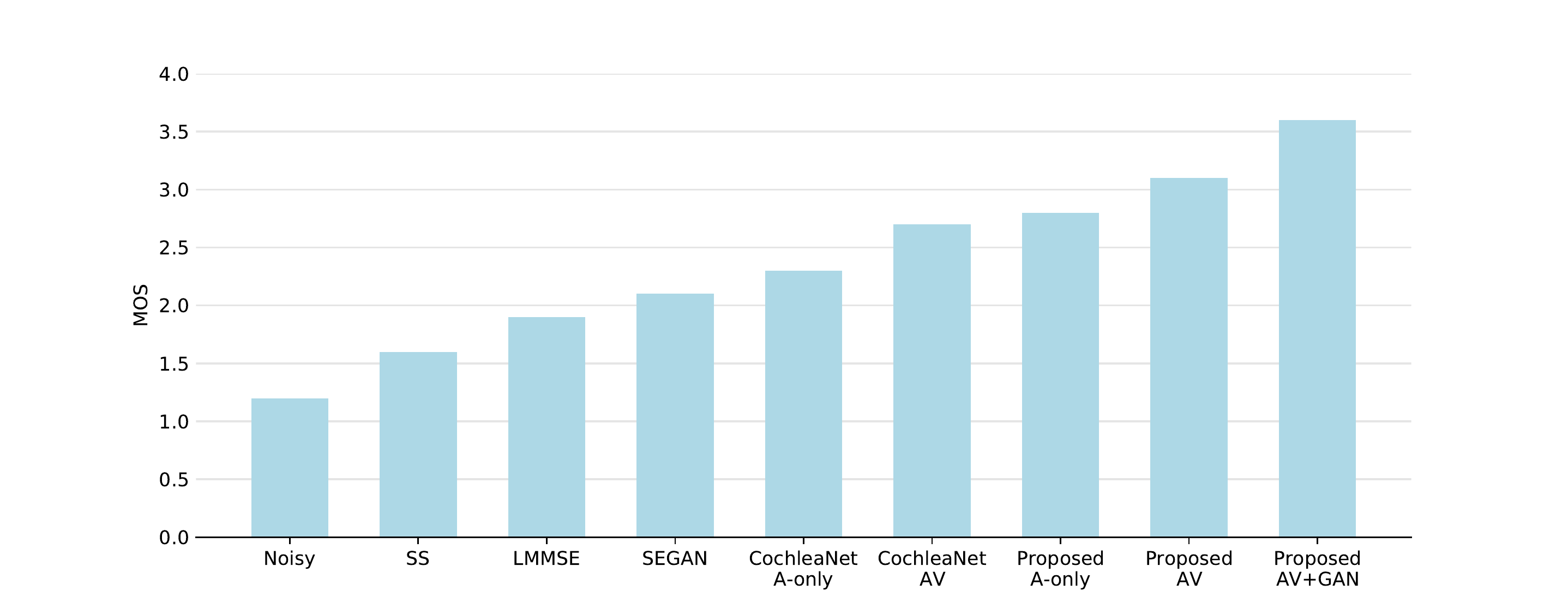}
    \caption{Subjective evaluation using mean-opinion score listening test for real noisy ASPIRE and VISION corpora}
    \label{fig:mushra_score}
\end{figure*}

The proposed AV model outperform SS~\cite{boll1979spectral},  linear minimum mean square error~\cite{ephraim1985speech}, SEGAN+~\cite{pascual2017segan} A-only CochleaNet, and AV CochleaNet. It can be seen that AV CochleaNet performs similar to A-only CochleaNet in the presence of visual noise. However, the proposed AV model is less affected by the added visual imperfection. 
In the low SNR ranges from (i.e. $SNR < 0$  dB), the AV outperforms the A-only model. On the other hand, AV model obtained STOI score of 0.59 (-12 dB), 0.63 (-9 dB) and 0.66 (-6 dB), as compared to 0.55 (-12 dB), 0.60 (-9 dB) and 0.62 (-6 dB) STOI score obtained by A-only model for speaker independent test set.
However, at high SNRs (i.e. $SNR >= 0$  dB) proposed AV model perform similar to A-only model, where AV model achieved STOI score of 0.69 (0 dB), 0.74 (3 dB), and 0.76 (6 dB), as compared to  0.65 (0 dB), 0.69 (3 dB) and 0.74 (0 dB) achieved by A-only model for speaker independent test set. In addition, it has been shown that the A-only and AV model significantly outperform A-only and AV CochleaNet.

\subsubsection{Scale-Invariant Signal-to-Distortion Ratio (SI-SDR) comparison} 

SI-SDR is modified scale invariant version of SDR. SI-SDR is used to predict the distortion introduced by the separated signal and usually defined the ratio between clean signal and distortion energy. The higher SDR shows the better quality of SE and less distortion in the enhanced speech. Fig.~\ref{fig:SISDR_gc3_GAN},~\ref{fig:SISDR_tcd_GAN},~\ref{fig:SISDR_hou_GAN} depicts the SI-SDR for proposed A-only and AV model, A-only CochleaNet, AV CochleaNet, SEGAN, SS, and linear minimum mean square error for speaker independent, large vocabulary and language independent test set. 

The proposed A-only and AV models outperform SS~\cite{boll1979spectral}, linear minimum mean square error~\cite{ephraim1985speech}, SEGAN+~\cite{pascual2017segan} A-only CochleaNet, and AV CochleaNet~\cite{gogate2020cochleanet}. In addition, the AV CochleaNet performs similar to the A-only CochleaNet in the presence of visual noise. The proposed AV model is less affected by the added visual imperfection. 
In addition, for the lower SNR ($SNR < 0$ dB) the proposed AV outperforms A-only model especially, where AV model achieved the SI-SDR score of 3.50 (-12 dB), 4.45 (-9 dB), and 5.04 (-6dB) as compared to 2.53 (-12 dB), 4.39 (-9 dB), and 5.62 (-6 dB) SI-SDR score obtained by A-only model for speaker independent test set.
However, AV perform similar to A-only model at high SNRs levels (i.e. $SNR >= 0$  dB), where AV model achieved SI-SDR score of 7.35 (0 dB), 8.01 (3 dB) and 8.35 (6dB), as compared to  7.25, 7.97 and 8.35 achieved by A-only model for speaker independent test set. It is worth to mention that, the proposed A-only and AV model perform better than A-only and AV CochleaNet. 

\begin{figure*}[!t]
    \centering
    \includegraphics[width=\linewidth]{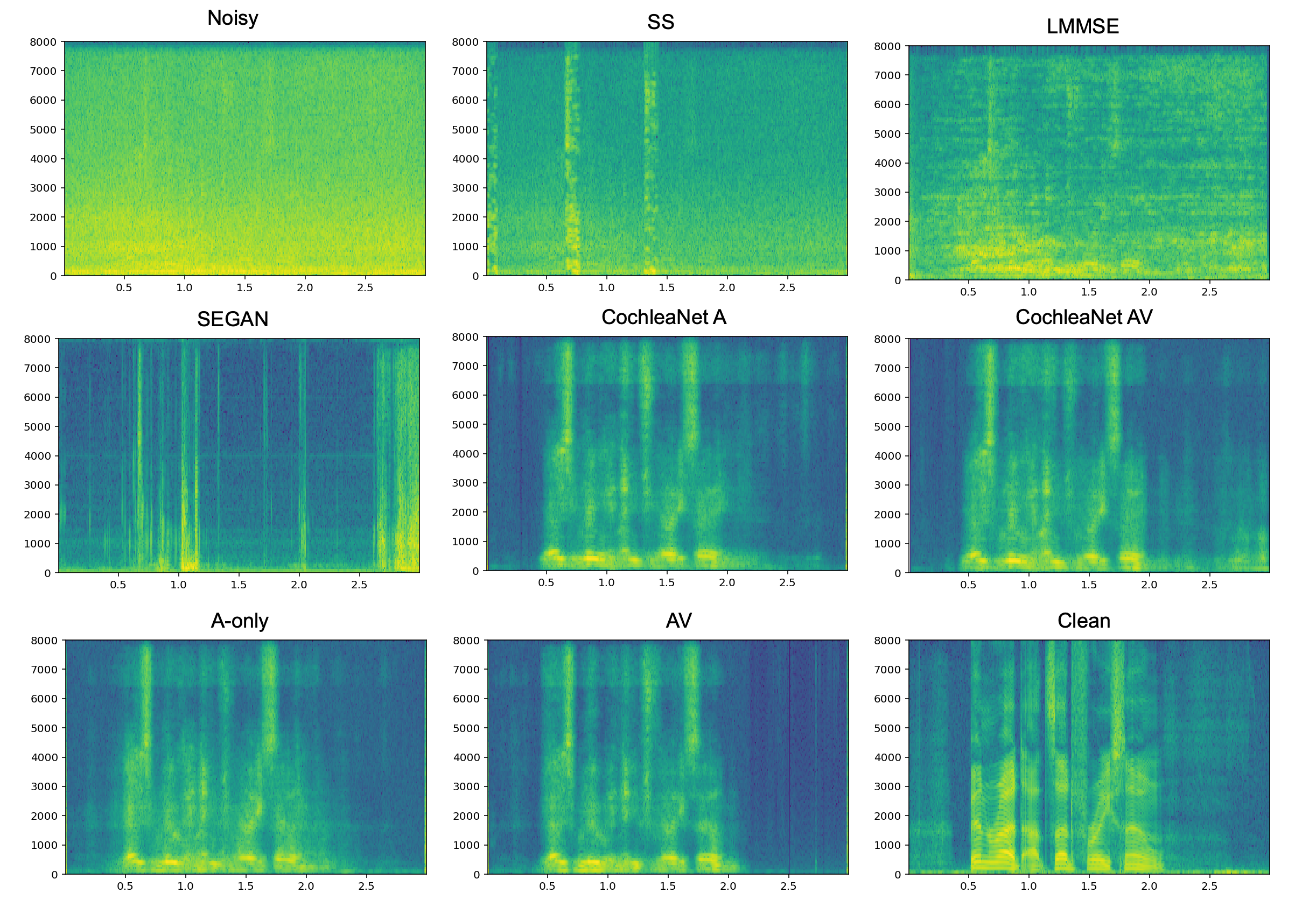}
    \caption{Spectogram comparison of a randomly selected utterance at -12 dB SNR from GRID-CHiMe3 corpus}
    \label{fig:spec}
\end{figure*}

Fig.~\ref{fig:spec} illustrates the spectogram for the resynthesised speech signal of a randomly selected utterance from GRID and Chime 3 AV corpus using proposed A-only and AV models as well as state-of-the-art approaches such as SS, LMMSE, SEGAN+, A-only and AV CochleaNet. In addition, spectogram for clean and noisy speech signal is shown for comparison. It is to be noted that, the speech is completely swamped with street noise and the performance of the proposed model is closer to clean spectogram. The state-of-the-art A-only approaches are unable to recover the speech components from the noisy signal. Finally, it can be seen the AV models are able to better reconstruct target speech as compared to A-only models specifically in silent speech regions.   

\subsection{Subjective listening tests on ASPIRE and VISION corpus}

The subjective speech quality can be computationally approximated using state-of-the-art objective evaluation metrics including PESQ and STOI~\cite{pesq, stoi}. However, human listening tests needs to be conducted to accurately understand subjective speech quality. As a result, we used the mean opinion score (MOS) type listening test for comparative subjective evaluation in our study. The data used for these tests include ASPIRE~\cite{gogate2020cochleanet} and VISION corpora~\cite{gogate2020vision} recorded in real noisy environment with a range of visual and acoustic noises. Twenty native English speakers (twelve men and eight women) with normal hearing volunteered to participate in the listening test. The listeners were first trained with five utterances and the purpose of the study was explained. In each individual listening test, twenty utterances chosen at random from the ASPIRE and VISION corpora were played. Listeners were asked to rate the enhanced speech quality on a scale of 0 to 5, with 0 representing incomprehensible, 1 representing very annoying, 2 representing annoying, 3 representing slightly annoying, 4 representing perceptible but annoying, and 5 representing perceptible.

SEGAN, SS, LMMSE, A-only and AV CochleaNet, proposed A-only and AV are comparatively evaluated along with noisy audio as reference. The speech quality scores for the aforementioned models are presented in Figure~\ref{fig:mushra_score}. It can be seen that, our proposed A-only and AV models significantly outperform SS, LMMSE, A-only and AV CochleaNet. It can be seen that, the proposed AV model can handle the visual imperfections present in VISION corpora as compared to state-of-the-art CochleaNet model. 

\subsection{Processing latency}

The processing latency for a SE algorithm can be defined as the difference between the time of arrival of speech and the time when the model finish processing the noisy speech. The latency is generally measured in milliseconds (ms). The ideal processing latency of a listening device depend on the severity of the hearing loss. The latency of most commercial hearing aids generally range from 8-12 ms. The processing latency of the proposed model is 20 ms. The latency is dependent upon the Fourier window shift (12 ms), STFT (0.5ms), ISTFT (0.5ms), and model prediction time (7 ms). These values are calculated with a 2.2 GHz Quad-Core Intel Core i7 with 16 GB RAM. The processing latency is mainly affected by the shift of Fourier window and the model processing time. The model processing time can be further optimised using mixed precision processing and quantization. In addition, the window shift can be decreased for exploiting the model in listening devices. Currently, the model can be used in real-time for video conferencing application.

\subsection{Limitations}

The limitation with the proposed framework is that (1) the resynthesised speech using IBM ignores phase estimation and result in invalid STFT problem (2) the model is trained for separating speech from non-speech noise and cannot separate multiple speakers (3) the model currently do not consider the noise level in the environment for contextual switching between A-only and AV model (4) the enhanced speech cannot be localised in the space as the model only supports processing single channel data.

\section{Conclusion} \label{sec:conclusiongan}
In this paper, we presented a novel framework for robust real-time AV SE that contextually takes account of both visual and acoustic noises. Specifically, a GAN architecture is developed to tackle the challenging issue of visual imperfections encountered in real noisy environments. Further, the GAN is integrated with a real-time AV SE model to contextually exploit noisy visual and acoustic speech independently of the SNR, in order to estimate a binary spectral mask for SE. The model is evaluated on benchmark visual speech noises from real world recordings as well as noisy speech recorded in the presence of multiple competing background sources. Preliminary performance evaluation in terms of objective metrics and subjective listening tests demonstrates significant improvement of our proposed AV SE framework compared to the state-of-the-art A-only (including SS, LMMSE) approaches as well as DNN based AV approaches (including benchmark SEGAN and CochleaNet models). Comparative experimental results indicate that our framework has the ability to work effectively in a range of SNRs with both visual and acoustic noise and can be deployed in real-world environments.
Ongoing work includes evaluation of the proposed framework in low-latency video conferencing applications. In future, we intend to optimise the framework using intelligibility-oriented loss functions and self-supervised learning. In addition, privacy concerns associated with the practical deployment of our AV SE framework will be explored. 
\printbibliography

@article{adeel2020contextual,
  title={Contextual deep learning-based audio-visual switching for speech enhancement in real-world environments},
  author={Adeel, Ahsan and Gogate, Mandar and Hussain, Amir},
  journal={Information Fusion},
  volume={59},
  pages={163--170},
  year={2020},
  publisher={Elsevier}
}

@inproceedings{gogate2020deep,
  title={Deep neural network driven binaural audio visual speech separation},
  author={Gogate, Mandar and Dashtipour, Kia and Bell, Peter and Hussain, Amir},
  booktitle={2020 International Joint Conference on Neural Networks (IJCNN)},
  pages={1--7},
  year={2020},
  organization={IEEE}
}

@article{gogate2020cochleanet,
  title={CochleaNet: A robust language-independent audio-visual model for real-time speech enhancement},
  author={Gogate, Mandar and Dashtipour, Kia and Adeel, Ahsan and Hussain, Amir},
  journal={Information Fusion},
  volume={63},
  pages={273--285},
  year={2020},
  publisher={Elsevier}
}

@article{adeel2019lip,
  title={Lip-reading driven deep learning approach for speech enhancement},
  author={Adeel, Ahsan and Gogate, Mandar and Hussain, Amir and Whitmer, William M},
  journal={IEEE Transactions on Emerging Topics in Computational Intelligence},
  year={2019},
  publisher={IEEE}
}

@inproceedings{gogate2017novel,
  title={A novel brain-inspired compression-based optimised multimodal fusion for emotion recognition},
  author={Gogate, Mandar and Adeel, Ahsan and Hussain, Amir},
  booktitle={2017 IEEE Symposium Series on Computational Intelligence (SSCI)},
  pages={1--7},
  organization={IEEE}
}

@inproceedings{gogate2018dnn,
  title={DNN Driven Speaker Independent Audio-Visual Mask Estimation for Speech Separation},
  author={Gogate, Mandar and Adeel, Ahsan and Marxer, Ricard and Barker, Jon and Hussain, Amir},
  booktitle={Interspeech 2018},
  pages={2723--2727},
  year={2018},
  organization={ISCA}
}

@inproceedings{gogate2020vision,
  title={Visual Speech In Real Noisy Environments (VISION): A Novel Benchmark Dataset and Deep Learning-Based Baseline System},
  author={Gogate, Mandar and Dashtipour, Kia and Hussain, Amir},
  booktitle={Interspeech 2020},
  pages={4521-4525},
  year={2020},
  organization={ISCA}
}

@inproceedings{adeel2017towards,
  title={Towards next-generation lipreading driven hearing-aids: A preliminary prototype demo},
  author={Adeel, Ahsan and Gogate, Mandar and Hussain, Amir},
  booktitle={Proceedings of the International Workshop on Challenges in Hearing Assistive Technology (CHAT-2017), Stockholm, Sweden},
  volume={19},
  year={2017}
}

@inproceedings{zhu2017unpaired,
  title={Unpaired image-to-image translation using cycle-consistent adversarial networks},
  author={Zhu, Jun-Yan and Park, Taesung and Isola, Phillip and Efros, Alexei A},
  booktitle={Proceedings of the IEEE international conference on computer vision},
  pages={2223--2232},
  year={2017}
}

@inproceedings{isola2017image,
  title={Image-to-image translation with conditional adversarial networks},
  author={Isola, Phillip and Zhu, Jun-Yan and Zhou, Tinghui and Efros, Alexei A},
  booktitle={Proceedings of the IEEE conference on computer vision and pattern recognition},
  pages={1125--1134},
  year={2017}
}

@article{targ2016resnet,
  title={Resnet in resnet: Generalizing residual architectures},
  author={Targ, Sasha and Almeida, Diogo and Lyman, Kevin},
  journal={arXiv preprint arXiv:1603.08029},
  year={2016}
}

@InProceedings{chung2016,
  author       = "Chung, J.~S. and Zisserman, A.",
  title        = "Lip Reading in the Wild",
  booktitle    = "Asian Conference on Computer Vision",
  year         = "2016",
}

@article{world2021hearing,
  title={Hearing screening: considerations for implementation},
  author={World Health Organization and others},
  year={2021},
  publisher={World Health Organization}
}

@inproceedings{sisdr,
  title={SDR--half-baked or well done?},
  author={Le Roux, Jonathan and Wisdom, Scott and Erdogan, Hakan and Hershey, John R},
  booktitle={ICASSP 2019-2019 IEEE International Conference on Acoustics, Speech and Signal Processing (ICASSP)},
  pages={626--630},
  year={2019},
  organization={IEEE}
}

@incollection{newmarch2017ffmpeg,
  title={Ffmpeg/libav},
  author={Newmarch, Jan},
  booktitle={Linux Sound Programming},
  pages={227--234},
  year={2017},
  publisher={Springer}
}

@inproceedings{donahue2018exploring,
  title={Exploring speech enhancement with generative adversarial networks for robust speech recognition},
  author={Donahue, Chris and Li, Bo and Prabhavalkar, Rohit},
  booktitle={2018 IEEE International Conference on Acoustics, Speech and Signal Processing (ICASSP)},
  pages={5024--5028},
  year={2018},
  organization={IEEE}
}

@article{lesica2018hearing,
  title={Why do hearing aids fail to restore normal auditory perception?},
  author={Lesica, Nicholas A},
  journal={Trends in neurosciences},
  volume={41},
  number={4},
  pages={174--185},
  year={2018},
  publisher={Elsevier}
}

@InProceedings{martinez2020lipreading,
  author       = "Martinez, Brais and Ma, Pingchuan and Petridis, Stavros and Pantic, Maja",
  title        = "Lipreading using Temporal Convolutional Networks",
  booktitle    = "ICASSP",
  year         = "2020",
}

@article{Grid,
	title        = {An audio-visual corpus for speech perception and automatic speech recognition},
	author       = {Cooke, Martin and Barker, Jon and Cunningham, Stuart and Shao, Xu},
	year         = 2006,
	journal      = {The Journal of the Acoustical Society of America},
	publisher    = {ASA},
	volume       = 120,
	number       = 5,
	pages        = {2421--2424}
}

@inproceedings{CHiME3,
	title        = {The third ‘CHiME’speech separation and recognition challenge: Dataset, task and baselines},
	author       = {Barker, Jon and Marxer, Ricard and Vincent, Emmanuel and Watanabe, Shinji},
	year         = 2015,
	booktitle    = {Automatic Speech Recognition and Understanding (ASRU), 2015 IEEE Workshop on},
	pages        = {504--511},
	organization = {IEEE}
}

@article{TCDTIMIT,
	title        = {TCD-TIMIT: An Audio-Visual Corpus of Continuous Speech},
	author       = {N. {Harte} and E. {Gillen}},
	year         = 2015,
	month        = may,
	journal      = {IEEE Transactions on Multimedia},
	volume       = 17,
	number       = 5,
	pages        = {603--615},
	doi          = {10.1109/TMM.2015.2407694},
	issn         = {1520-9210}
}

@article{snyder2015musan,
	title        = {Musan: A music, speech, and noise corpus},
	author       = {Snyder, David and Chen, Guoguo and Povey, Daniel},
	year         = 2015,
	journal      = {arXiv preprint arXiv:1510.08484}
}

@article{NOISEX92,
	title        = {Assessment for automatic speech recognition: II. NOISEX-92: A database and an experiment to study the effect of additive noise on speech recognition systems},
	author       = {Varga, Andrew and Steeneken, Herman JM},
	year         = 1993,
	journal      = {Speech communication},
	publisher    = {Elsevier},
	volume       = 12,
	number       = 3,
	pages        = {247--251}
}

@article{hou2018audio,
	title        = {Audio-visual speech enhancement using multimodal deep convolutional neural networks},
	author       = {Hou, Jen-Cheng and Wang, Syu-Siang and Lai, Ying-Hui and Tsao, Yu and Chang, Hsiu-Wen and Wang, Hsin-Min},
	year         = 2018,
	journal      = {IEEE Transactions on Emerging Topics in Computational Intelligence},
	publisher    = {IEEE},
	volume       = 2,
	number       = 2,
	pages        = {117--128}
}

@article{dlib09,
	title        = {Dlib-ml: A Machine Learning Toolkit},
	author       = {Davis E. King},
	year         = 2009,
	journal      = {Journal of Machine Learning Research},
	volume       = 10,
	pages        = {1755--1758}
}

@inproceedings{pesq,
	title        = {Perceptual evaluation of speech quality (PESQ)-a new method for speech quality assessment of telephone networks and codecs},
	author       = {Rix, Antony W and Beerends, John G and Hollier, Michael P and Hekstra, Andries P},
	year         = 2001,
	booktitle    = {2001 IEEE International Conference on Acoustics, Speech, and Signal Processing. Proceedings (Cat. No. 01CH37221)},
	volume       = 2,
	pages        = {749--752},
	organization = {IEEE}
}

@article{stoi,
	title        = {An algorithm for intelligibility prediction of time--frequency weighted noisy speech},
	author       = {Taal, Cees H and Hendriks, Richard C and Heusdens, Richard and Jensen, Jesper},
	year         = 2011,
	journal      = {IEEE Transactions on Audio, Speech, and Language Processing},
	publisher    = {IEEE},
	volume       = 19,
	number       = 7,
	pages        = {2125--2136}
}

@article{ephraim1985speech,
	title        = {Speech enhancement using a minimum mean-square error log-spectral amplitude estimator},
	author       = {Ephraim, Yariv and Malah, David},
	year         = 1985,
	journal      = {IEEE transactions on acoustics, speech, and signal processing},
	publisher    = {IEEE},
	volume       = 33,
	number       = 2,
	pages        = {443--445}
}

@inproceedings{boll1979spectral,
	title        = {A spectral subtraction algorithm for suppression of acoustic noise in speech},
	author       = {Boll, S},
	year         = 1979,
	booktitle    = {Acoustics, Speech, and Signal Processing, IEEE International Conference on ICASSP'79.},
	volume       = 4,
	pages        = {200--203},
	organization = {IEEE}
}

@article{hu2007evaluation,
	title        = {Evaluation of objective quality measures for speech enhancement},
	author       = {Hu, Yi and Loizou, Philipos C},
	year         = 2007,
	journal      = {IEEE Transactions on audio, speech, and language processing},
	publisher    = {IEEE},
	volume       = 16,
	number       = 1,
	pages        = {229--238}
}

@article{pascual2017segan,
	title        = {SEGAN: Speech Enhancement Generative Adversarial Network},
	author       = {Pascual, Santiago and Bonafonte, Antonio and Serr{\`a}, Joan},
	year         = 2017,
	journal      = {Proc. Interspeech 2017},
	pages        = {3642--3646}
}

@article{lu2019audio,
  title={Audio--visual deep clustering for speech separation},
  author={Lu, Rui and Duan, Zhiyao and Zhang, Changshui},
  journal={IEEE/ACM Transactions on Audio, Speech, and Language Processing},
  volume={27},
  number={11},
  pages={1697--1712},
  year={2019},
  publisher={IEEE}
}

@article{michelsanti2019deep,
  title={Deep-learning-based audio-visual speech enhancement in presence of Lombard effect},
  author={Michelsanti, Daniel and Tan, Zheng-Hua and Sigurdsson, Sigurdur and Jensen, Jesper},
  journal={Speech Communication},
  volume={115},
  pages={38--50},
  year={2019},
  publisher={Elsevier}
}

@article{montesinos2021cappella,
  title={A cappella: Audio-visual Singing Voice Separation},
  author={Montesinos, Juan F and Kadandale, Venkatesh S and Haro, Gloria},
  journal={arXiv preprint arXiv:2104.09946},
  year={2021}
}

@article{gao2021visualvoice,
  title={VisualVoice: Audio-Visual Speech Separation with Cross-Modal Consistency},
  author={Gao, Ruohan and Grauman, Kristen},
  journal={arXiv preprint arXiv:2101.03149},
  year={2021}
}

@article{kim2021multi,
  title={A Multi-Resolution Approach to GAN-Based Speech Enhancement},
  author={Kim, Hyung Yong and Yoon, Ji Won and Cheon, Sung Jun and Kang, Woo Hyun and Kim, Nam Soo},
  journal={Applied Sciences},
  volume={11},
  number={2},
  pages={721},
  year={2021},
  publisher={Multidisciplinary Digital Publishing Institute}
}

@article{afouras2018conversation,
  title={The conversation: Deep audio-visual speech enhancement},
  author={Afouras, Triantafyllos and Chung, Joon Son and Zisserman, Andrew},
  journal={arXiv preprint arXiv:1804.04121},
  year={2018}
}

@article{afouras2019my,
  title={My lips are concealed: Audio-visual speech enhancement through obstructions},
  author={Afouras, Triantafyllos and Chung, Joon Son and Zisserman, Andrew},
  journal={arXiv preprint arXiv:1907.04975},
  year={2019}
}

@inproceedings{arriandiaga2021audio,
  title={Audio-visual target speaker enhancement on multi-talker environment using event-driven cameras},
  author={Arriandiaga, Ander and Morrone, Giovanni and Pasa, Luca and Badino, Leonardo and Bartolozzi, Chiara},
  booktitle={2021 IEEE International Symposium on Circuits and Systems (ISCAS)},
  pages={1--5},
  year={2021},
  organization={IEEE}
}

@article{aldeneh2020self,
  title={Self-supervised learning of visual speech features with audiovisual speech enhancement},
  author={Aldeneh, Zakaria and Kumar, Anushree Prasanna and Theobald, Barry-John and Marchi, Erik and Kajarekar, Sachin and Naik, Devang and Abdelaziz, Ahmed Hussen},
  journal={arXiv preprint arXiv:2004.12031},
  year={2020}
}

@article{abel2014novel,
  title={Novel two-stage audiovisual speech filtering in noisy environments},
  author={Abel, Andrew and Hussain, Amir},
  journal={Cognitive Computation},
  volume={6},
  number={2},
  pages={200--217},
  year={2014},
  publisher={Springer}
}

\end{document}